# Strong influence of precursor powder on the critical current density of Fe-sheathed MgB$_2$ tapes

Dongliang Wang[1], Yanwei Ma[1,#], Zhengguang Yu[1], Zhaoshun Gao[1], Xianping Zhang[1], K. Watanabe[2], and E. Mossang[3]

[1] Applied Superconductivity Lab., Institute of Electrical Engineering, Chinese Academy of Sciences, P. O. Box 2703, Beijing 100080, China

[2] Institute for Materials Research, Tohoku University, Sendai 980-8577, Japan

[3] Grenoble High Magnetic Field Laboratory, C.N.R.S., 25, Avenue des Martyrs, B.P. 166, 38 042 Grenoble Cedex 9, France

E-mail: ywma@mail.iee.ac.cn

**Abstract:**

The effect of the quality of starting powders on the microstructure and superconducting properties of in-situ processed Fe-sheathed MgB$_2$ tapes has been investigated. Three different types of commercial atomized spherical magnesium powder and two different purities of amorphous boron powder were employed. When using the 10-micrometre magnesium as precursor powders, the Mg reacted with boron more uniformly and quickly, thus the uniformity of the fabricated MgB$_2$ was improved and the grain size of the MgB$_2$ was decreased, hence significant critical current density (Jc) enhancements were achieved for MgB$_2$ tapes. Jc at 4.2 K for MgB$_2$ tapes made from the 10 μm Mg and high purity boron powders was at least a factor of ten higher than values measured for MgB$_2$ samples made from all other starting powders. At 20 K, 5 T, the typical Jc values of the tapes were over $1.0 \times 10^4$ A/cm$^2$ and were much better than those of tape samples reported recently.

---

[#] Author to whom correspondence should be addressed

# 1. Introduction

Since the discovery of the superconductivity of $MgB_2$ in 2001, great interest was aroused in the scientific community. That is because $MgB_2$ has several advantages, such as high transition temperature (39 K), low cost and weak-link free grain coupling and so on, compared to conventional low-temperature superconductors. So far the most popular process of $MgB_2$ tape and wire fabrication is the so-called in situ powder-in-tube (PIT) process [1], however, the critical current density Jc values of in situ-processed $MgB_2$ tapes are still below the practical level. In order to improve the property, research efforts have been directed towards either enhancing critical current density (Jc) by improving grain connectivity, or improving in-field performance through introducing pinning centers, such as chemical doping [2-6], irradiation with heavy ions [7], and various processing techniques [8-9].

On the other hand, the quality of the starting Mg and B powders, such as purity and particle size, has been demonstrated to play an important role in determining the superconducting properties of $MgB_2$. It has been reported that Mg and B mixture powder prepared by ball milling is effective to improve the grain connections and enhance the Jc values [10-11]. Several groups have suggested that the purity of boron powder has a noticeable effect on Jc [12-15]. At the same time, it has also been found that the replacement of commercial Mg powder with $MgH_2$ powder [16] and nanometer-size Mg powder [17] is effective to enhance the reaction between Mg and B, thus a better grain connectivity and hence the Jc enhancement. However, the $MgH_2$ and nanometer-size Mg powders reported previously are definitely expensive, compared to the commercial Mg powder. If the cheap, finer Mg powder can be used to make $MgB_2$, the material cost could be decreased significantly.

In this work, different cheap atomized spherical magnesium is used as precursor powders for the fabrication of Fe-sheathed $MgB_2$ tapes by the PIT technique. The influence of Mg and B starting powders on the microstructure and superconducting properties of tapes will be discussed in detail by employing x-ray diffraction, scanning electron microscope and transport measurements.

## 2. Experiment details

MgB$_2$ tapes were prepared with Mg and B powders by the in-situ PIT process. Commercial atomized spherical magnesium powders with three different sizes (10 μm, 20 μm and 70 μm, 99.8%, Tangshan Weihao Co.) and two different purities of amorphous boron powders (97% and 99.99%, Zhongpu Ruituo Co.) were used as starting powders. The magnesium powder and boron powder were mixed in the atom ratio of 1:2, followed by hand milling using an agate mortar and pestle for about 1 h. The mixed powders were put into pure Fe tubes with 8 mm in outer diameter and 5 mm in inner diameter. The tubes were subsequently rotate swaged, drawn and rolled into tapes. The final size of the tapes was about 3.8 mm in width and 0.5 mm in thickness. Finally, short samples of about 40 mm wrapped in Ta foils were heat-treated at 700°C or 800°C for 1 h, and then followed by cooling down to room temperature in the furnace. The argon gas was allowed to flow into the furnace during the heat-treatment process to reduce the oxidation of the samples. Depending on the precursor powder used, the tape samples were labeled as M10, M20, M70, and M10-B99, respectively, as shown in table I.

Phase identification was performed by x-ray diffraction (XRD) using Cu Kα radiation. For study of tapes, the Fe sheaths were mechanically removed by peeling off the Fe sheath to expose the core. Microstructural observation was carried out by scanning electron microscopy (SEM). DC magnetization measurement was performed with a superconducting quantum interference device (SQUID) magnetometer. The Tc value was determined by taking the first deviation point from linearity that signifies the transition from the normal to superconducting state. Transport critical current (Ic) of tape samples were measured at 4.2 and 20 K and its magnetic field dependence were evaluated by using a standard four-probe technique. The criterion for the Ic definition was 1μV/cm. A magnetic field was applied parallel to the tape surface. For each set of tape made from different types of precursor powders, the Ic measurement was performed for several samples to check reproducibility.

## 3. Results and discussion

Figure 1 shows x-ray diffraction patterns of the superconducting cores of MgB$_2$

tapes sintered at 700°C fabricated using different size starting Mg powders. $MgB_2$ is found to be the main phase in all samples, with minor impurity phases of MgO present. Clearly, no remnant Mg is observed for the M10 and M10-B99 samples, indicating that the boron powder has reacted completely with Mg. It is noted that the amount of MgO in the M10-B99 sample is smaller than that in the tape M10. However, for the M20 and M70 samples, remnant Mg and impurity phases of $Fe_2B$ are also observed, suggesting that the reaction of large size Mg with B is not complete, and some boron would react with Fe to form $Fe_2B$.

Figure 2 shows the superconducting transition temperatures for all the tape samples heat-treated at 700°C determined by DC susceptibility. M10-B99 tapes show a strong sharp transition, suggesting a fairly strong coupling of grains. The Tc obtained as the onset of magnetic screening is about 36.8K. Meanwhile, other tapes made from different size Mg and 97% boron powers exhibit almost the same Tc of 35 K, that is slightly lower than that of M10-B99 samples. This lowering of the Tc can be attributed to the higher oxygen mole fraction in it. Further, the transition width of the 97% purity boron samples is larger than that of the 99.99% purity boron sample, suggesting large cores inhomogeneity for the low purity boron samples. It is noted that our results are in good agreement with those of Ribeiro et al who found that Tc was improved with increasing the boron purity [14].

Figure 3(a) shows the transport critical current density Jc at 4.2 K in magnetic fields up to 12 T for tapes with different size Mg powders sintered at 700°C. Only data above 4 T are shown, because at lower field region, Ic was too high to be measured. The striking result of Fig. 4(a) is that both tapes made from 10 μm Mg powder present much higher values of Jc than the samples made from 20 μm and 70 μm Mg powders. When we compare the M10 and M10-B99 tapes, we should note the difference of the Jc-B property. The Jc value of M10-B99 tapes was higher than that of M10 tape in the high magnetic fields, suggesting that high purity boron powder (99.99%) is more effective in improving the in-field Jc than the low-grade amorphous powder (97%), very consistent with previous report [13, 15]. Clearly, the largest Jc values were achieved in the M10-B99 samples, more than an order of magnitude higher than all other tapes in high field regions. At 4.2 K, the Jc reached up to 10 kA/cm$^2$ in 9 T and was still about 1.7 kA/cm$^2$ in 12 T. This Jc increase is due to the improved flux pinning by using finer Mg powder and high purity boron powder as will be discussed below. In addition, what is more interesting is the result for the M70 sample: its critical current density is better than that of the M20 samples. The reason is probably that the grain homogeneity of the M20 is not as good as that of the M70 sample. A comparison with other Jc values is presented in Fig. 3, also showing about twice as large as those of tapes prepared from the commercial Mg (Alfa, 325 mesh) and 99.99% purity boron powders [18]. These results clearly demonstrate that the 10 μm Mg powder is much effective to enhance the Jc values of MgB$_2$ tapes.

Figure 3(b) shows the Jc-H curves at 4.2 K of three MgB$_2$ tapes sintered at different temperatures. Jc decreases, and the magnetic field dependence of Jc increases with increasing the heat-treatment temperature, meaning that the reduction of Jc with increasing the heat-treatment temperature can be explained by the decrease of irreversibility fields [19].

Figure 4 shows the transport Jc-H curves at 20 K for the M10-B99 tapes heat-treated at 700°C. The Jc at 20 K decreased with increasing the field, and the Jc values higher than 10 kA/cm$^2$ were obtained in a field of 5 T for M10-B99 samples. It should be noted that this Jc value is even higher than the tapes fabricated with nanometer-size Mg prepared by the thermal plasma method [20], and highlights the importance of atomized spherical magnesium powders used in this work for enhancing the Jc of MgB$_2$ superconductors at high temperatures.

Microstructural analyses are employed to further elucidate the mechanism for the size effect of Mg powder on Jc of MgB$_2$ tapes. Figure 5 shows typical SEM images of the fractured MgB$_2$ cores for all the tapes sintered at 700°C. The MgB$_2$ cores for all the tapes show a similar microstructure; however, the morphology of the M10-B99 sample was refined to smaller and more homogeneous grains compared to other MgB$_2$ ones. The average grain sizes of the M10, M20, M70 and M10-B99 MgB$_2$ samples are about 230nm, 270nm, 320nm and 200nm, respectively. The grain size of the M10-B99 tape is smaller than that of the tapes prepared with low purity boron powders. On the other hand, the average grain size of the tapes prepared with 97% boron powders becomes smaller with reducing the particle size of magnesium powders. Actually, the grain refinement with the 10 μm Mg powder is further supported by the full width at half maximum (FWHM) results for the $MgB_2$ (002) peak, as shown in Fig. 6. Clearly, the FWHM value for the M10-B99 sample is the largest, suggesting that the crystallite size of $MgB_2$ for the M10-B99 sample is smallest among the samples investigated. Therefore, it can be explained by that through using finer magnesium powder and high purity boron powder, the reaction of Mg and B occurs more uniformly and quickly; thus the uniformity of the fabricated $MgB_2$ core is improved and finer grain size is obtained. Obviously, the SEM investigation shows that decreasing the particle size of magnesium powder and increasing the purity of boron powder are two effective means to improve the core quality of the fabricated $MgB_2$.

The high performance of Jc in high magnetic fields in the $MgB_2$ tapes by using the 10 μm Mg powders as precursor powder may be attributed to strong grain boundary pinning in the samples. As revealed by microstructural analyses, the grain size decreases with reducing the particle size of Mg starting powders. Thus the enhanced number of grain boundaries associated with the smaller grain size can increase the flux pinning, like in the case of high-Jc, high-$H_{c2}$ $Nb_3Sn$ superconductors [21]. Furthermore, 10 μm-size fine Mg powders have higher reactivity, and the reaction of Mg and B occurs more uniformly and rapidly; hence, the fast reaction between Mg and B suppresses the formation of $Fe_2B$, and hence a high Jc was achieved. On the contrary, the lower Jc values of M20 and M70 samples are closely related to the suppressed $MgB_2$ phase formation. This is because the remnant Mg and the formation of $Fe_2B$ were clearly existed, as supported by the XRD analyses shown in figure 1. The formation of $Fe_2B$ means the deviation of the starting stoichiometric composition of Mg + B as $MgB_2$. At the same time, high-temperature heat treatment causes the vaporization of the unreacted Mg, resulting in a decrease in the density of the core layer. Therefore, a reduction in Jc was observed.

**4. Conclusions**

MgB$_2$/Fe tapes were fabricated with different size magnesium and different purity boron powders by the in-situ powder-in-tube process. It is found that the size of staring Mg and the purity of boron powders is important in determining the Jc(H) performance of MgB$_2$ tapes: the smaller the starting Mg powders, the better the Jc(H) performance. When using 10 μm size Mg powder as a precursor, the Mg can react with B powder more uniformly and quickly, thus the uniformity of the fabricated MgB$_2$ was improved and the grain size of the MgB$_2$ was decreased. The Jc values of tapes with 10 μm Mg and high purity B powders were about one order of magnitude larger than those of tapes made from other size Mg starting powders and reached 1667 A/cm$^{-2}$ in a field of 12 T at 4.2 K. The Jc of 1.0x10$^4$ A/cm$^2$ at 20 K, 5 T of the tapes is the highest value ever reported for tapes prepared by the in situ process. Our results suggest that in order to get high performance MgB$_2$/Fe tapes, it is quite important to use the finer size Mg powders as well as high purity starting boron powders.

**Acknowledgments**

The authors thank Jinming Liu, Liye Xiao, G. Nishijima, S. Awaji and L. Z. Lin for their help and useful discussion. This work is partially supported by the National Science Foundation of China under Grant Nos. 50472063 and 50572104, National '973' Program (Grant No. 2006CB601004) and National '863' Project (Grant No. 2006AA03Z203). Also, this work at the GHMFL has been supported by the "Transnational Access to Infrastructures - Specific Support Action" Program (Contract No. RITA-CT-2003-505474 of the European Commission).

TABLE I.  The starting powders of the fabricated tapes

| Tape sample | Size of magnesium powder (μm) | Purity of boron powder |
|:---:|:---:|:---:|
| M10 | 10 | 97% |
| M20 | 20 | 97% |
| M70 | 70 | 97% |
| M10-B99 | 10 | 99.99% |

# Captions

Figure 1　XRD patterns of in situ processed all the tapes made from different size Mg powders sintered at 700°C. The data were obtained after peeling off the Fe-sheath.

Figure 2　Normalized magnetic susceptibility against temperature for all $MgB_2$ tapes heat-treated at 700°C. The inset shows an enlarged view near the superconducting transitions.

Figure 3　Transport critical current densities at 4.2 K as a function of magnetic field for $MgB_2$/Fe tapes sintered at different temperatures: (a) 700°C and (b) 800°C. The measurements were performed in magnetic fields parallel to the tapes surface.

Figure 4　Transport critical current densities versus applied magnetic field at 20 K for $MgB_2$/Fe tapes prepared with the 10 μm size Mg and high purity boron powders. The heat-treatment temperature was 700°C.

Figure 5 SEM images of the fractured $MgB_2$ core layers of Fe-sheathed tapes: (a) M10; (b) M20; (c) M70; (d) M10-B99 after heat treatment at 700°C.

Figure 6 The relationship between the samples heat-treated at 700℃ and the FWHM values of the $MgB_2$ (002) peak. The inset shows the 002 XRD peak profiles of the four tapes.

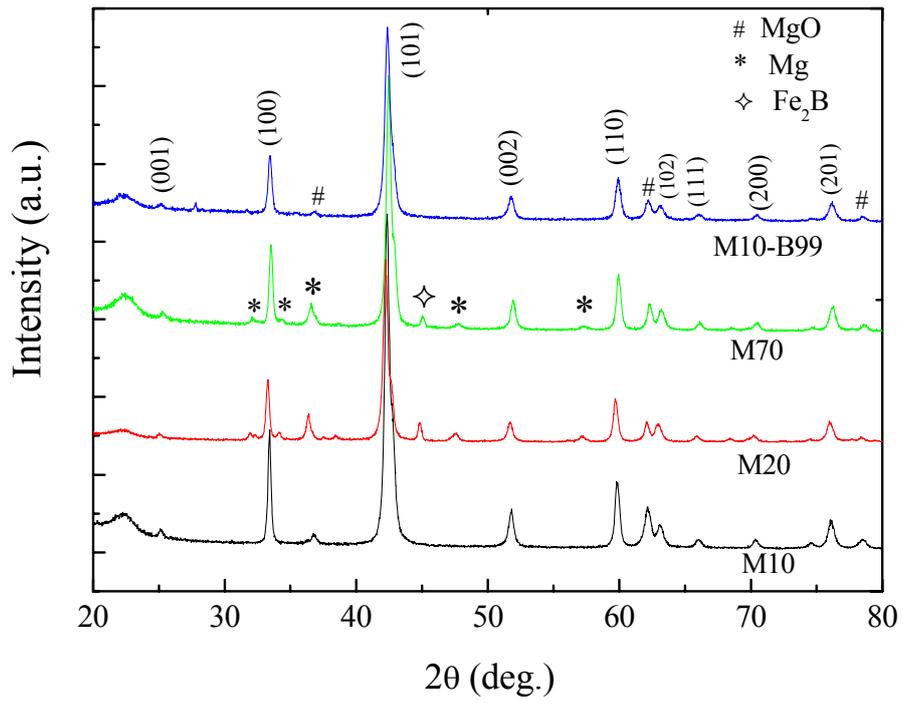

Fig.1 Wang et al.

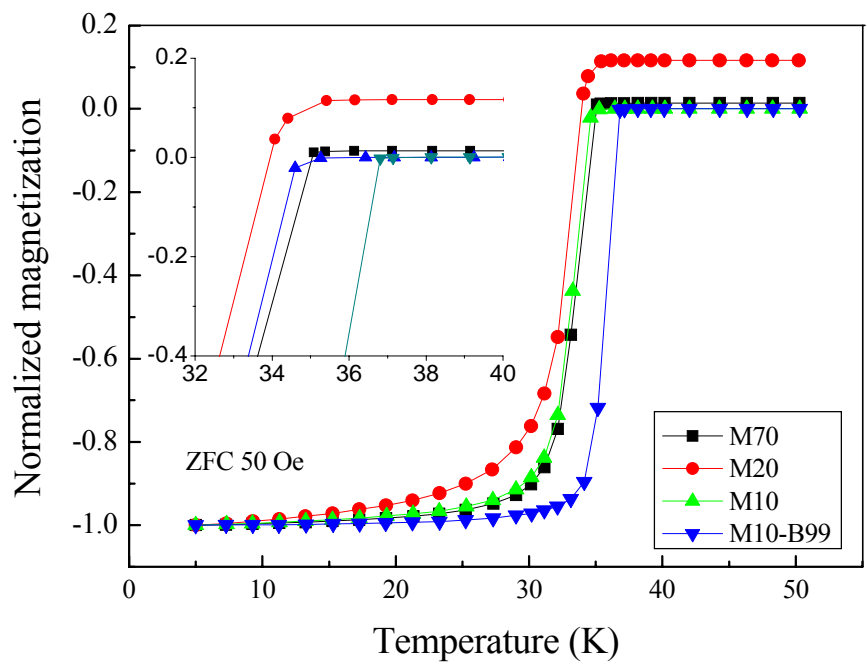

Fig.2 Wang et al.

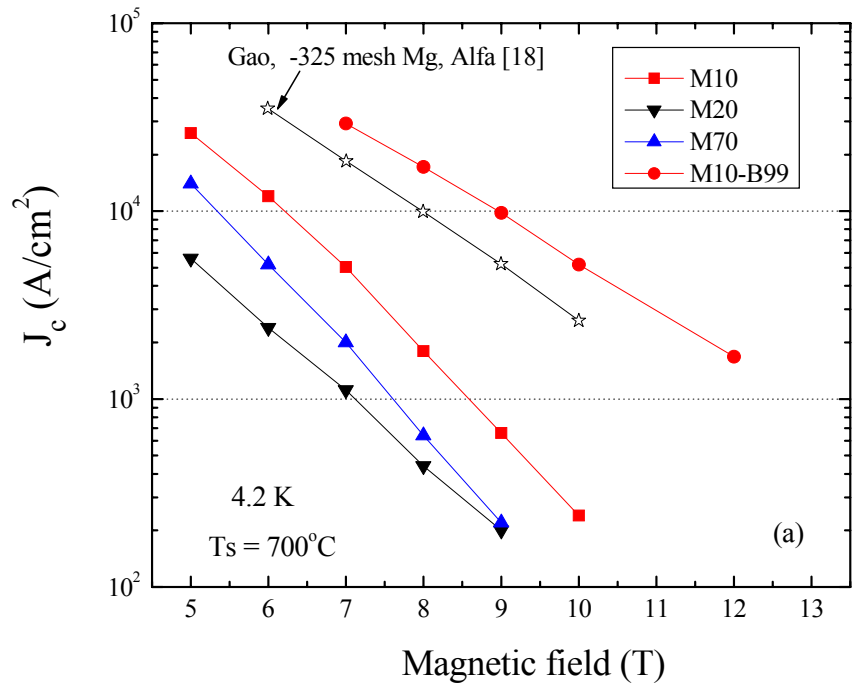

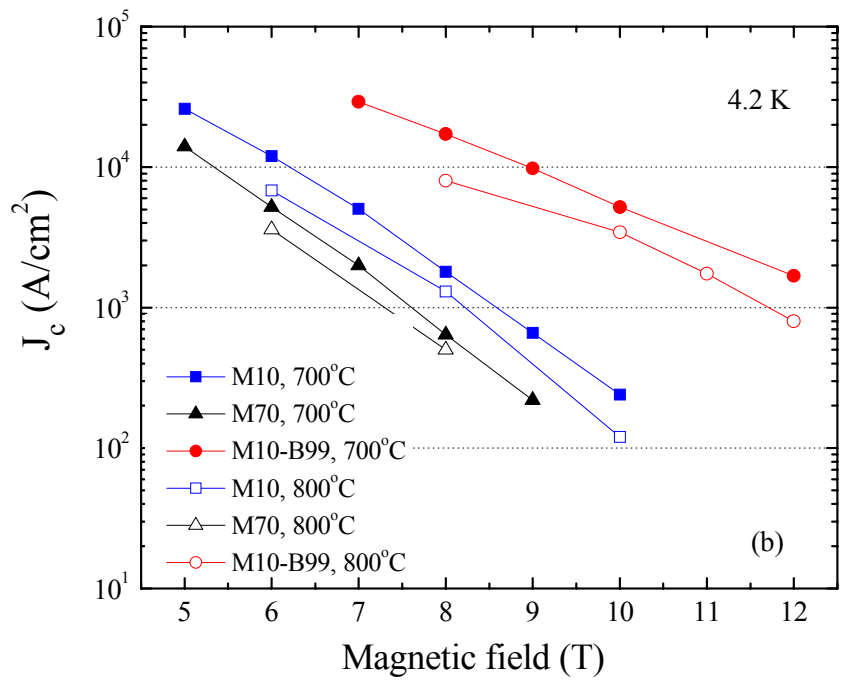

Fig.3 Wang et al.

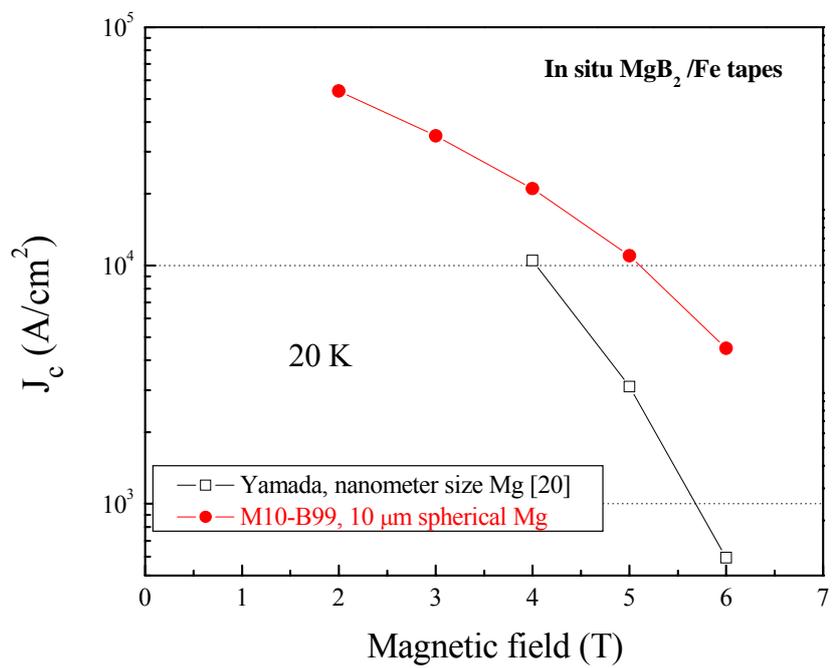

Fig.4 Wang et al.

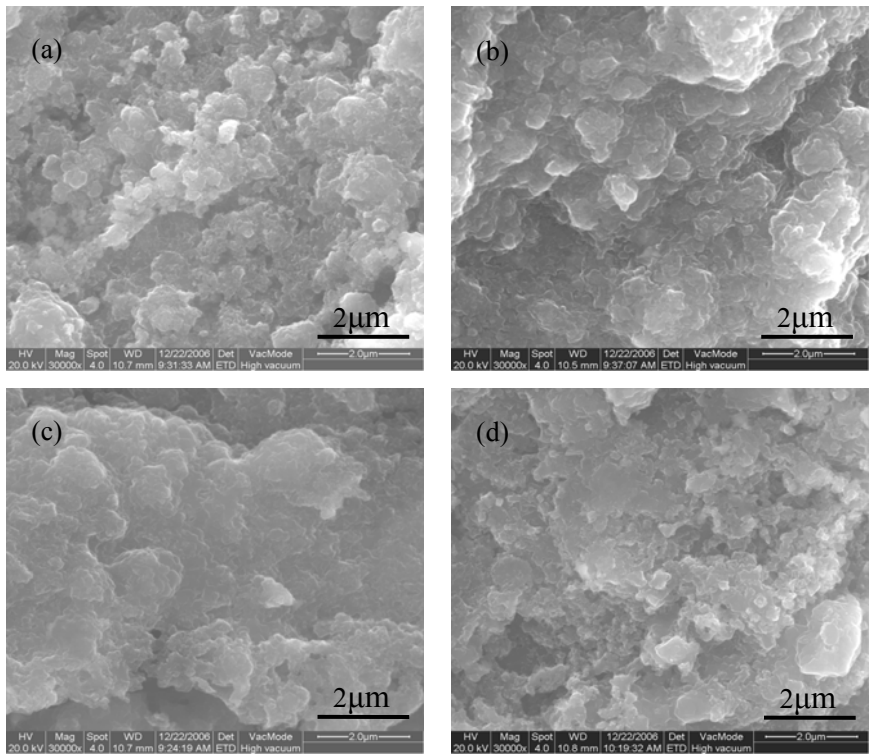

Fig.5 Wang et al.

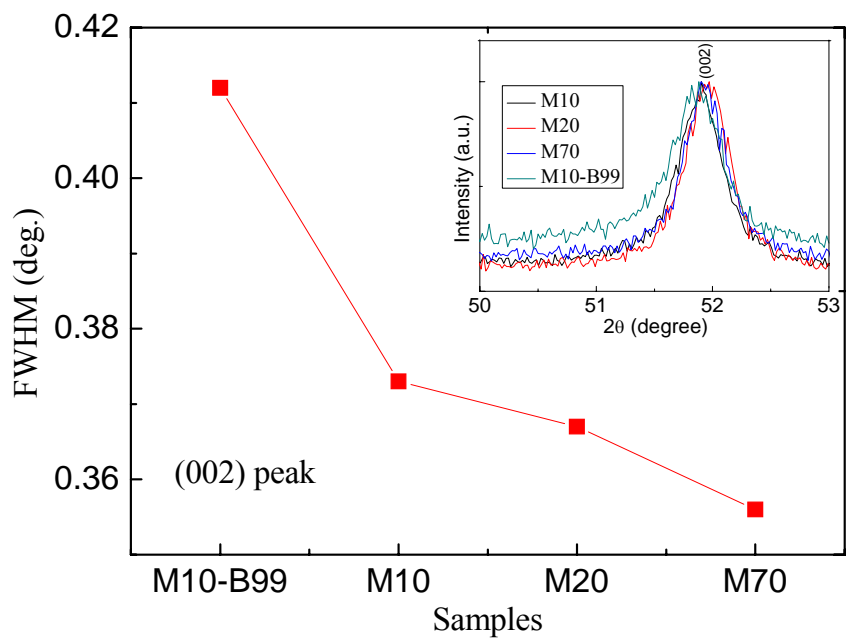

Fig.6 Wang et al.